\documentclass[%
reprint,
superscriptaddress,
nofootinbib,
amsmath,amssymb,
aps,
pra,
]{revtex4-1}

\usepackage{graphicx}
\usepackage{dcolumn}% Align table columns on decimal point
\usepackage{grffile}
\usepackage{dcolumn}
\usepackage{bm}
\usepackage{epsfig}
\usepackage{mathrsfs}  %  package for the "curly" fonts 
\usepackage{subfigure}
\usepackage{multirow}
\usepackage{epstopdf}
\usepackage{amsmath}
\usepackage{algorithmicx}
\usepackage{amssymb}
\usepackage{tensor}
\usepackage[math]{cellspace}
\usepackage{bookmark}
\usepackage[usenames,dvipsnames]{xcolor}
\usepackage{hyperref}
\usepackage{slashed}
\usepackage{youngtab}
\usepackage{physics}
\usepackage{tensor}
\usepackage[capitalise]{cleveref}
\usepackage{orcidlink}

\crefformat{equation}{Eq.~(#2#1#3)}
\crefformat{table}{Tab.~(#2#1#3)}
\crefformat{figure}{Fig.~(#2#1#3)}
\crefformat{appendix}{App.~(#2#1#3)}
\crefformat{section}{Sec.~(#2#1#3)}
\crefformat{subsection}{Subsec.~(#2#1#3)}

\makeatletter
\renewcommand*\env@matrix[1][\arraystretch]{%
  \edef\arraystretch{#1}%
  \hskip -\arraycolsep
  \let\@ifnextchar\new@ifnextchar
  \array{*\c@MaxMatrixCols c}}
\makeatother

\allowdisplaybreaks
\begin{document}

\title{Black-hole binaries and waveforms in Quadratic Gravity
}

\author{Aaron Held\,\orcidlink{0000-0003-2701-9361}}
\email{aaron.held@phys.ens.fr}
\affiliation{
Institut de Physique Théorique Philippe Meyer, Laboratoire de Physique de l’\'Ecole normale sup\'erieure (ENS), Universit\'e PSL, CNRS, Sorbonne Universit\'e, Universit\'e Paris Cité, F-75005 Paris, France
}

\author{Hyun Lim \orcidlink{0000-0002-8435-9533}}
\email{hyunlim@lanl.gov}
\thanks{\\Both authors contributed equally. The names are listed alphabetically.}
\affiliation{
Applied Computer Science (CCS-7), 
Los Alamos National Laboratory, 
Los Alamos, NM  87545 USA
}
\affiliation{Center for Theoretical Astrophysics,
Los Alamos National Laboratory,
Los Alamos, NM 87545 USA 
}

\begin{abstract}
We report on the first numerical-relativity simulations of black-hole binaries that deviate from General Relativity due to quadratic-curvature corrections. Said theory of Quadratic Gravity propagates additional massive modes and admits both Kerr and non-Kerr black-hole solutions. We chose the respective masses ``at threshold'', i.e., such that (at least) one of the black holes dynamically transitions from the Kerr to the non-Kerr branch during the early inspiral. The subsequent waveforms differ from their General Relativity counterparts throughout inspiral, merger, and ringdown.
\end{abstract}

%\pacs{}
\maketitle

%%%%%%%%%%%%%%%%%%%%%%%%%%%%%%%%%%%%%%%%%%%%%%%%%%%%%%%%%
%%%%%%%%%%%%%%%%%%%%%%%%%%%%%%%%%%%%%%%%%%%%%%%%%%%%%%%%%
\noindent{\textbf{Introduction.}}
%%%%%%%%%%%%%%%%%%%%%%%%%%%%%%%%%%%%%%%%%%%%%%%%%%%%%%%%%
%%%%%%%%%%%%%%%%%%%%%%%%%%%%%%%%%%%%%%%%%%%%%%%%%%%%%%%%%
The age of gravitational-wave astronomy is in full swing. Following the first detection of a gravitational wave~\cite{LIGOScientific:2016aoc}, the Ligo-Virgo-Kagra collaboration (LVK) now routinely detects binary events~\cite{LIGOScientific:2018mvr,LIGOScientific:2020ibl,KAGRA:2021vkt}. 
The resulting wealth of observational data provides the exciting opportunity to test General Relativity (GR) in the fully dynamical and nonlinear regime and at an unprecedented level of precision. 
Realizing this prospect relies on sufficiently accurate predictions for gravitational waveforms, both, within GR itself \emph{and} when corrections are included in the gravitational dynamics. 

Here, we focus on potential corrections to the gravitational action which occur at quadratic order in curvature, i.e., on Quadratic Gravity (QG)~\cite{Stelle:1976gc,Stelle:1977ry}. 
Corrections of cubic and quartic order have been considered as well, both, analytically~\cite{PhysRevD.94.104005,Endlich:2017tqa,Cano:2019ore,Sennett:2019bpc,deRham:2020ejn} and numerically~\cite{Cayuso:2020lca,Cayuso:2023aht}. Non-minimal couplings to other fields, e.g., to scalars, also occur at quadratic order in curvature, see~\cite{Kanti:1995vq,Alexander:2009tp}.
Field redefinitions can mix between different sectors and also between different orders of such a curvature expansion~\cite{Burgess:2003jk,Endlich:2017tqa}. Whenever said field redefinitions do not impact physical conclusions, any choice of frame should lead to the same physical outcome. In our choice of frame, the matter sector remains minimally coupled.
The respective Lagrangian of vacuum QG (neglecting a cosmological constant, see also~\cite{Buoninfante:2023ryt}) reads
\begin{align}
    \mathcal{L} = M_\text{Pl}^2\Bigg[
	\frac{1}{2}R
	+\frac{1}{12m_0^2}R^2
	-\frac{1}{4m_2^2}C_{\mu\nu\alpha\beta}C^{\mu\nu\alpha\beta}
\Bigg]\;,
    \label{eq:lagrangian-QG}
\end{align}
where $m_0$ and $m_2$ correspond to the masses of the additional massive spin-0 and spin-2 modes associated with the respective higher curvature corrections.
We set the speed of light to $c=1$ and express all observables in units of the Planck mass $M_\text{Pl} = 1/\sqrt{8\pi\,G}$ or equivalently in units of the Newton constant $G$. The vacuum dynamics of the theory is then determined by the dimensionless mass ratios $m_2/M_\text{Pl}$ and $m_0/M_\text{Pl}$ and the intrinsic scales $M_i$ of the initial data, e.g., the mass ratios $M_1/M_\text{Pl}$ and $M_2/M_\text{Pl}$ of the two black holes in a binary system. (In distinction, we denote the total mass~\cite{Arnowitt:1960es} of the binary by $M$.) The dynamics of vacuum GR is recovered if $m_0/M_i\gg 1$ and $m_2/M_i\gg 1$ (with $i=1,2$), see~\cite{Held:2023aap} for nonlinear simulations in this regime.

\begin{figure}[t]
    \centering
    \includegraphics[trim={0cm 0cm 0cm 0cm},clip,width=\linewidth]{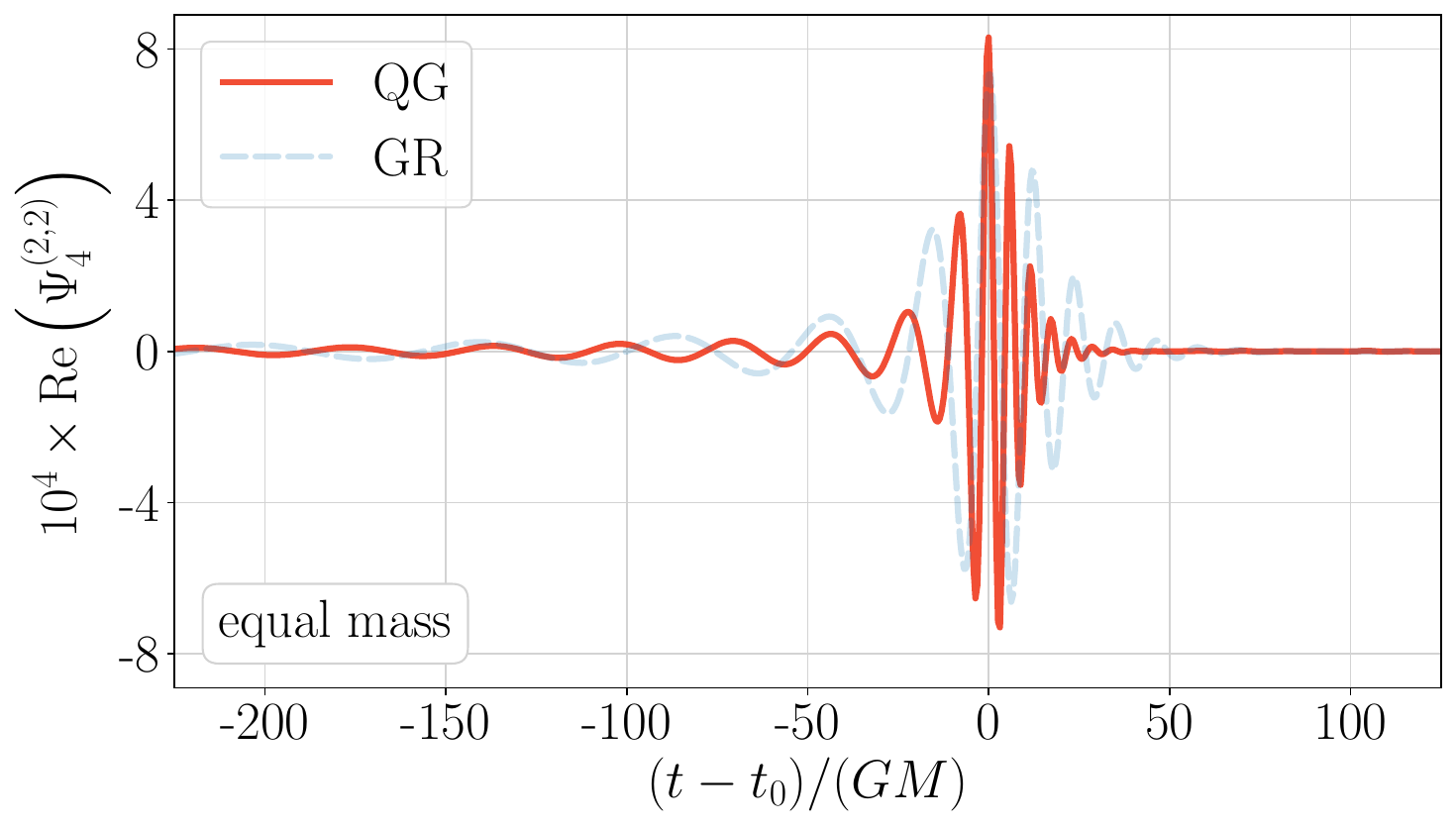}
    \includegraphics[trim={0cm 0cm 0cm 0cm},clip,width=\linewidth]{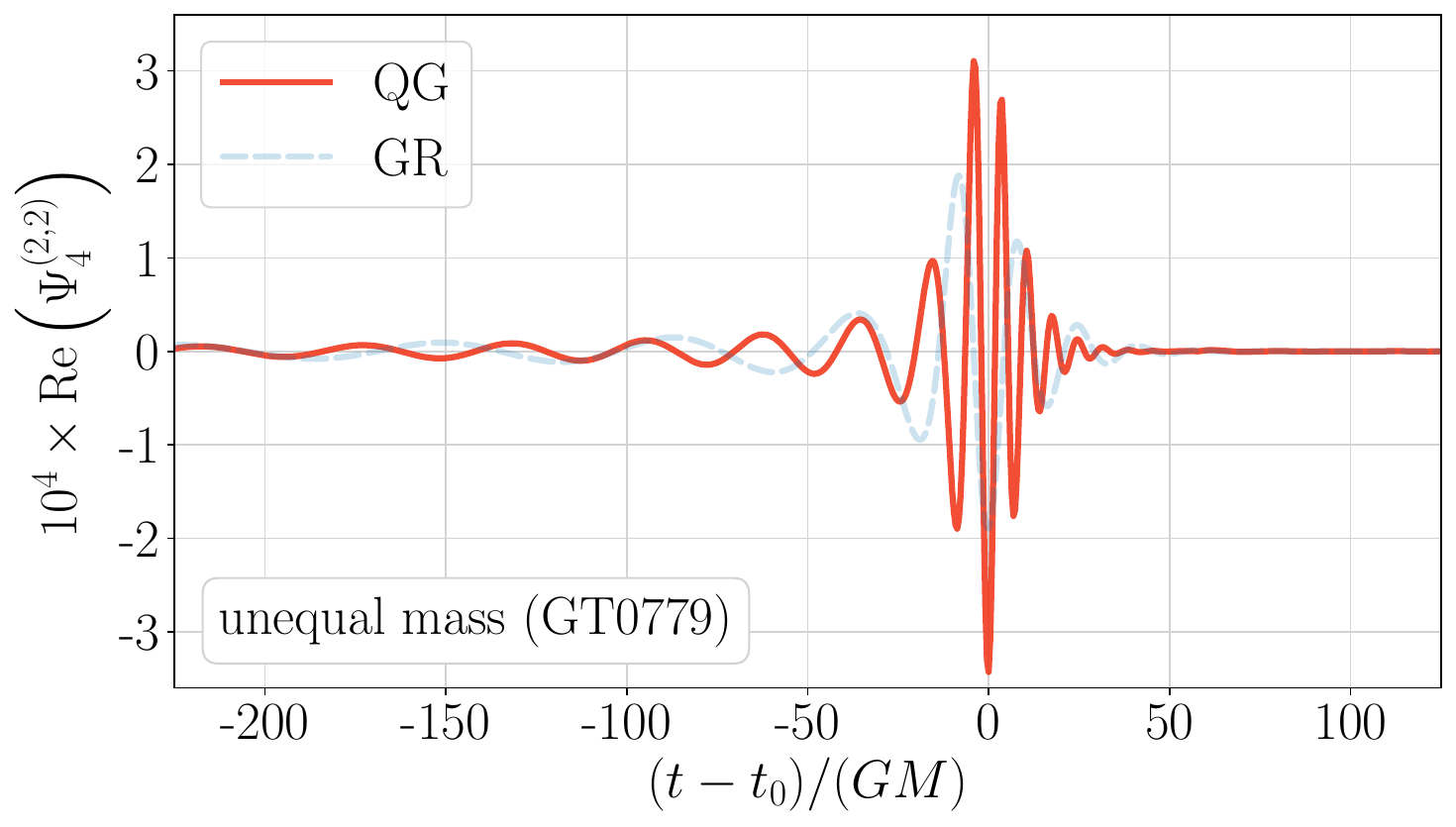}
    \caption{
    \label{fig:waveforms}
    Gravitational waveforms for an equal-mass (top) and an unequal-mass (bottom) binary merger, see~\cref{tab:binary-params} for the initial binary parameters and QG masses. We have aligned all waveforms at merger times $t_0$, i.e., at the peak absolute value of the waveform. Time is measured in units of the total initial mass $M$.
    The full waveforms, including the unphysical early-insprial phase and corresponding transitions of GR initial data, are shown in the supplementary material.
    }
\end{figure}

We focus on vacuum gravity because most of the observed gravitational-wave events have likely originated from sufficiently isolated black-hole binaries. Moreover, black holes provide an exceptionally clean probe of GR (see~\cite{2005GReGr..37.2253J,1923rmp..book.....B,Robinson:1975bv,1995CQGra..12..149K} for uniqueness theorems in GR). 
For the first time, we obtain full binary mergers and gravitational waveforms, see~\cref{fig:waveforms}, which are distinct from the predictions of GR because of the presence of the quadratic-curvature terms in~\cref{eq:lagrangian-QG}. Our numerical treatment~\cite{Held:2021pht,Held:2023aap} (detailed below) allows us to simulate the nonlinear theory as if it were a fundamental theory of nature. We comment on ghost instabilities and the interpretation of our results in the context of effective field theory (EFT) in the discussion.
\\

%

%%%%%%%%%%%%%%%%%%%%%%%%%%%%%%%%%%%%%%%%%%%%%%%%%%%%%%%%%
%%%%%%%%%%%%%%%%%%%%%%%%%%%%%%%%%%%%%%%%%%%%%%%%%%%%%%%%%
\noindent{\textbf{Numerical evolution.}}
%%%%%%%%%%%%%%%%%%%%%%%%%%%%%%%%%%%%%%%%%%%%%%%%%%%%%%%%%
%%%%%%%%%%%%%%%%%%%%%%%%%%%%%%%%%%%%%%%%%%%%%%%%%%%%%%%%%
Obtaining accurate and complete waveforms requires numerical relativity (NR): While NR simulations are computationally expensive, there is, at least at present, no other way to treat the nonlinear merger phase.
Most NR simulations are based on a (3+1) split of the metric into a spatial metric $\gamma_{ij}$ and a timelike unit normal vector $n^a$, i.e., in terms of the line element, $ds^2 = -\alpha^2\,dt^2+\gamma_{ij}(dx^i+\beta^i\,dt)(dx^j+\beta^j\,dt)$, with $\alpha$ and $\beta^i$ referred to as the lapse function and the shift vector. From the action, one can then derive the $(3+1)$ Hamiltonian, the equations of motion, and a constrained initial-value problem (IVP) for the respective canonical variables~\cite{Arnowitt:1960es}.
With suitable gauge and variable choices, the IVP for GR is (locally) well-posed~\cite{choquet2008general} and thus admits numerical treatment via discretization~\cite{Pretorius:2005gq,Baumgarte:2010ndz}. 
Convergence rates to the continuum limit are understood and under control~\cite{Calabrese2006,Babiuc2008,Giannakopoulos2020}.
With sufficient computing resources, the exact continuum solution can be approximated arbitrarily well and we thus refer to such formulations as ``numerical evolution without approximation''. 

To characterize any deviation from GR, the first nontrivial task is to find a well-posed initial-value formulation for the corresponding dynamics. Much attention has been devoted to non-minimally coupled scalar fields~\cite{Witek2019,Okounkova2019,Okounkova2020,Witek2020,Ripley2020a,Ripley2020b,Okounkova:2020rqw,East2021,Silva:2020omi,Figueras:2020dzx,Doneva:2024ntw} for which (local) well-posedness has been established at sufficiently weak non-minimal coupling~\cite{Kovacs:2020pns,Kovacs:2020ywu}.
Various approaches have been explored, including (i)
``iterative order-reduction methods'', which perturbatively truncate at the level of the field equations~\cite{Witek2019,Okounkova2019}, (ii)
``dissipative methods'' in which higher-order (Lorentz-violating) spatial derivatives are added~\cite{deRham:2023ngf} and, perhaps most promising, (iii) the ``fixing-the-equations approach''~\cite{Cayuso:2020lca,Cayuso:2023aht}, which introduces fiducial fields which, in turn, are dynamically dampened to their physical value~\cite{Cayuso:2017iqc}. It is a matter of current research to establish how far these approximation methods can accurately capture the nonlinear dynamics~\cite{GalvezGhersi:2021sxs,Franchini:2022ukz,Corman:2024cdr}.
For QG, we emphasize that no such approximation is necessary in the first place. Just as for GR, the initial-data evolution can be written in quasi-linear diagonal form~\cite{Noakes:1983} and a (locally) well-posed and numerically stable evolution scheme for QG has been established~\cite{Noakes:1983,Held:2021pht,Held:2023aap}. We use this scheme throughout the present work and refer to~\cite{Held:2023aap} for details. We also highlight that a recent well-posedness proof~\cite{Figueras:2024bba} suggests that the same methodology can be applied to a general class of gravitational actions.

The evolution system derived in~\cite{Held:2023aap} evolves the usual Baumgarte--Shapiro--Shibata--Nakamura (BSSN) variables~\cite{Shibata:1995we,Baumgarte:1998te}, i.e., the conformal factor $\phi$, the conformal (traceless) metric $\overline{\gamma}_{ij}$, as well as the trace and traceless part of the conformal extrinsic curvature, $A$ and $A_{ij}$, respectively. In vacuum GR, the above geometric evolution is closed by the Einstein equations which imply Ricci-flatness. In QG, the geometric evolution is closed by evolution equations for the Ricci-curvature variables, i.e., the Ricci scalar $\mathcal{R}$ and the traceless Ricci tensor $\widetilde{\mathcal{R}}_{ab}$.
The latter is decomposed into a spatial trace, a spatially traceless part, and a temporal part, $\mathcal{A}$, $\mathcal{A}_{ij}$, and $\mathcal{C}_i$, respectively, i.e., 
$
    \widetilde{\mathcal{R}}_{ab} = 
    \mathcal{A}_{ab} 
    + \frac{1}{3}\,\gamma_{ab}\,\mathcal{A}
    - 2\,n_{(a}\mathcal{C}_{b)}
    +n_an_b\,\mathcal{A}
 $.
 To achieve a first-order (in time) system, we also evolve the first-order counterparts of the above Ricci-curvature variables, see~\cite{Held:2023aap} for details. 
 We have implemented the evolution system in the \texttt{Dendro-GR}~\cite{Fernando2019} code framework and use the same numerical techniques described and benchmarked in~\cite{Held:2023aap} 
 \\

\begin{table}[t]
    {\renewcommand{\arraystretch}{1.5}
    \begin{tabular}{c||c|c||c|c|c|c}
        \hline\hline
        &
        \multicolumn{2}{c||}{QG masses} &
        \multicolumn{4}{c}{Binary parameters}
        \\\hline
        Case & 
        $GM_2m_0$ & $GM_2m_2$ & 
        $\sqrt{G}\,M_1$ & $q=\frac{M_1}{M_2}$ & $a_{z,1}$ & $a_{z,2}$
        \\\hline\hline
        equal mass &
        1 & 0.2 &
        1 & 1 & 0 & 0
        \\\hline
        GT0779 &
        1 & 0.2 &
        1 & 5 & -0.696 & 0
        \\\hline\hline
    \end{tabular}}
    \caption{
    \label{tab:binary-params} 
    Quadratic Gravity masses and binary parameters (in units of $G = 1/(8\pi\,M_\text{Pl}^2) = 1$), see also~\cite{Ferguson:2023vta}. We restrict to spin-aligned initial conditions (i.e., to $a_{x,1}=a_{y,1}=a_{x,2}=a_{y,2}=0$ in all cases) and choose initial momenta such as to achieve quasi-circular GR orbits~\cite{Jani:2016wkt}, see also~\cite{Husa:2007rh,Tichy:2010qa}.
    }
\end{table}
%

%%%%%%%%%%%%%%%%%%%%%%%%%%%%%%%%%%%%%%%%%%%%%%%%%%%%%%%%%
%%%%%%%%%%%%%%%%%%%%%%%%%%%%%%%%%%%%%%%%%%%%%%%%%%%%%%%%%
\noindent{\textbf{Dynamical departure from vacuum GR.
}}
%%%%%%%%%%%%%%%%%%%%%%%%%%%%%%%%%%%%%%%%%%%%%%%%%%%%%%%%%
%%%%%%%%%%%%%%%%%%%%%%%%%%%%%%%%%%%%%%%%%%%%%%%%%%%%%%%%%
First (3+1) simulations~\cite{Held:2023aap} suggest that QG can exactly mimic vacuum GR, even in the fully nonlinear strong-gravity regime, as long as $m_2$ is sufficiently heavy. Without matter sources, the only known dynamical departure from the Ricci-flat subsector occurs due to a well-established linear long-wavelength instability~\cite{Brito:2013wya,Lu:2017kzi,Collingbourne:2020sfy,Held:2022abx}. The latter affects only sufficiently small black holes below a threshold horizon radius of $r_g = 2\,GM \approx 0.87/m_2$, see~\cite{Brito:2013wya,East:2023nsk} for the impact of spin.

\begin{figure}[!t]
    \centering
    \includegraphics[trim={0cm 0cm 0cm 0cm},clip,width=\linewidth]{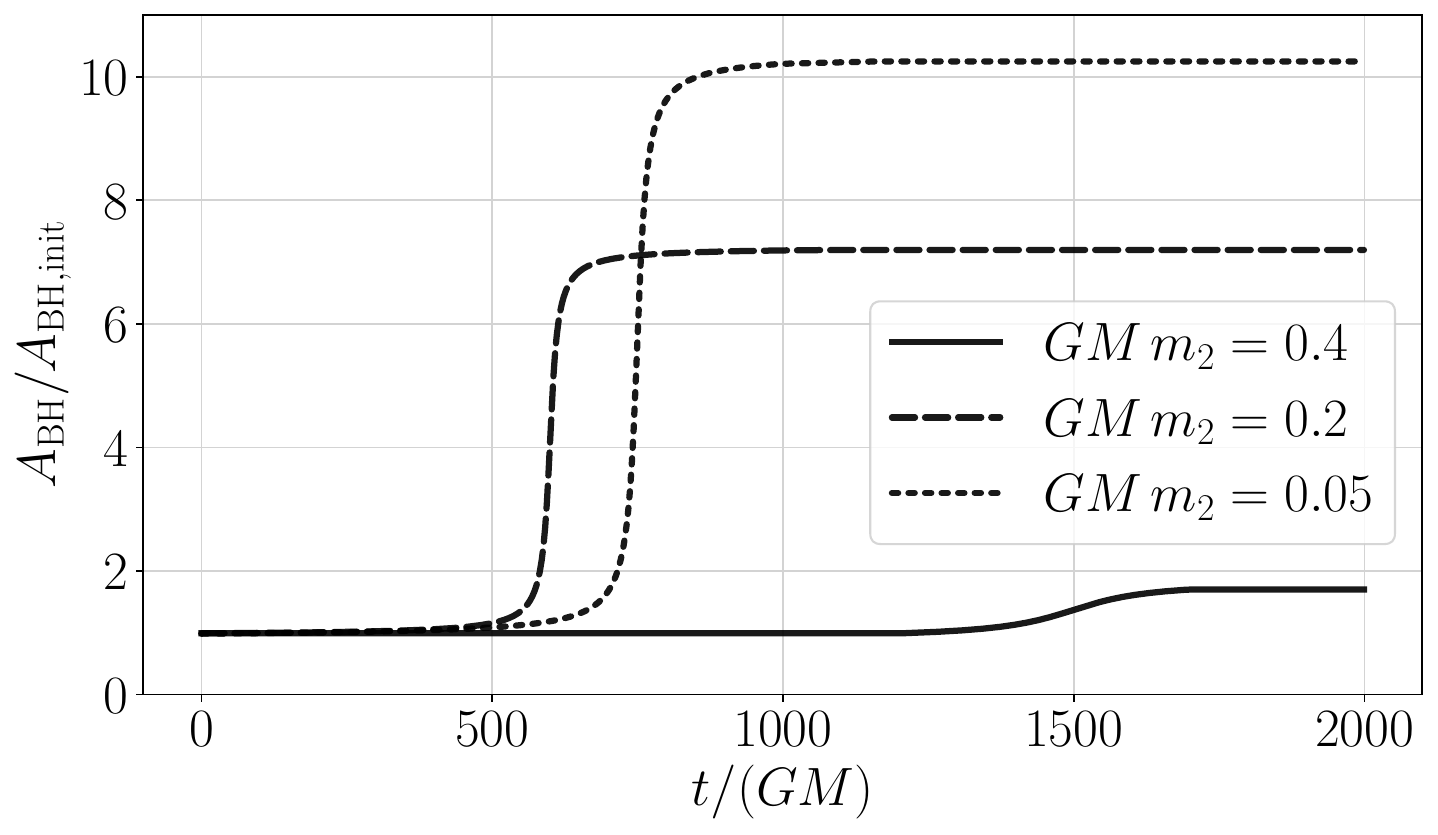}
    \caption{
    \label{fig:transitions}
    Transitions of Schwarzschild to non-Schwarzschild black holes as indicated by a growth of the apparent horizon~$A_\text{BH}$ (normalised to its initial value~$A_\text{BH,init}$). The onset of the transition is due to a linear long-wavelength instability and depends on the indicated value of $GM\,m_2$. (In all cases, we set $GM\,m_0=1$.) Nonlinearities quench the linear instability and the black hole settles in a distinct long-lived state.
    }
\end{figure}

In~\cref{fig:transitions}, we follow the resulting transitions between GR and non-GR black holes within the full (3+1) nonlinear evolution and for exemplary choices of $m_0$ and $m_2$, see also~\cite{East:2023nsk} for simulations in the $m_0/M\rightarrow \infty$ limit and the supplementary material for further comparison. 
The initial rate of instability matches expectations from the linear analysis~\cite{Brito:2013wya,Lu:2017kzi,Collingbourne:2020sfy,Held:2022abx}. Once nonlinearities become important, they quench the instability and the black hole settles in a new stable state --- or, at least, in a state that is long-lived compared to the simulation time. We have verified that the final state exhibits non-vanishing Ricci scalar curvature (see supplementary material).
\\

%

%%%%%%%%%%%%%%%%%%%%%%%%%%%%%%%%%%%%%%%%%%%%%%%%%%%%%%%%%
%%%%%%%%%%%%%%%%%%%%%%%%%%%%%%%%%%%%%%%%%%%%%%%%%%%%%%%%%
\noindent{\textbf{Initial data for binary systems ``at threshold''.}}
%%%%%%%%%%%%%%%%%%%%%%%%%%%%%%%%%%%%%%%%%%%%%%%%%%%%%%%%%
%%%%%%%%%%%%%%%%%%%%%%%%%%%%%%%%%%%%%%%%%%%%%%%%%%%%%%%%%
Given the above, we prepare initial data for binary systems of two rotating black holes in puncture gauge and choose the binary parameters and fundamental masses such that the lightest black hole(s) transition(s) from the GR to the non-GR branch during the ``early inspiral'' phase, see supplementary material for the full waveforms.
In~\cref{tab:binary-params} we provide the specific initial binary parameters. The latter are chosen such that we can compare the GR waveforms with the \texttt{MAYA} catalog~\cite{Ferguson:2023vta}. This benchmark test for our evolution code is presented in the supplementary material.
Facilitating the dynamical departure from GR (see~\cref{fig:transitions}), we can circumvent the difficulty of constructing consistent initial data for binary systems involving non-GR black holes\footnote{Constructing initial data for binary systems beyond GR is challenging and has not yet been attempted in QG, see, e.g.,~\cite{Brady:2023dgu,Nee:2024bur} for recent progress in scalar-tensor theories.}. 
To ensure that the respective transitions occur sufficiently early in the inspiral, we fix $m_2$ such that the transition time for the lower-mass binary companion is minimal, i.e., we fix $GM_2\,m_2 = 0.2$.
In this case~\cite{Held:2022abx}, the timescale of the linear instability is given by $t_\text{GL} \approx 20\,GM_2$, i.e., every $t_\text{GL}$, the linear instability grows one order in magnitude. We add Gaussian random noise with a magnitude of $10^{-7}$ to the initial data and, hence, anticipate that the transitions are triggered by the linear instability and occur within the first $t_\text{transition} \approx 100-200\,GM_2$ of the evolution. As long as $t_\text{transition}\ll t_\text{merger}$, the merger waveform is thus asymptotically close to that of a binary system in which the transition to the non-GR branch occurred already at asymptotically large binary separation. 
In~\cref{fig:waveforms}, we only show this ``post-transition'' portion of the waveform.

Regarding the comparison to the respective GR waveform, we expect that the transition modifies the binary parameters. This includes the local mass (i.e., the surface of the apparent horizon) as well as the asymptotic mass of the respective binary companion. We note that beyond GR the local and asymptotic mass need no longer be the same.
We also expect a subsequent modification of the relative radial velocity in the binary. Hence, the equivalent asymptotic (in time) initial data may correspond to an eccentric beyond-GR binary, even if, as is the case for our simulations, the respective GR binary is quasi-circular. 
A GR waveform with suitably adapted masses and eccentricity may thus be more closely aligned with the QG waveform. Due to the involved nonlinearities, however, we do not expect that adapting the binary parameters in GR can fully mimic the QG waveform.
We will address the question of waveform degeneracy in dedicated future work. 
\\

%%%%%%%%%%%%%%%%%%%%%%%%%%%%%%%%%%%%%%%%%%%%%%%%%%%%%%%%%
%%%%%%%%%%%%%%%%%%%%%%%%%%%%%%%%%%%%%%%%%%%%%%%%%%%%%%%%%
\noindent{\textbf{Expected observational constraints.}}
%%%%%%%%%%%%%%%%%%%%%%%%%%%%%%%%%%%%%%%%%%%%%%%%%%%%%%%%%
%%%%%%%%%%%%%%%%%%%%%%%%%%%%%%%%%%%%%%%%%%%%%%%%%%%%%%%%%
Our choice of $m_2/M_\text{Pl}$ (in relation to the smaller black-hole mass $M_2/M_\text{Pl}$) is ``at threshold'', i.e., minimizes the transition time. For larger $m_2/M_\text{Pl}$, the instability eventually shuts off completely and QG mimics GR without any deviations~\cite{Held:2023aap}. For smaller $m_2/M_\text{Pl}$, the transition time is prolonged but the instability remains present. While the latter means that the black holes transition more slowly, cf.~\cref{fig:transitions}, the dynamical end-point, i.e., the respective non-Kerr black hole, exhibits increasingly large deviations from Kerr spacetime. Hence, we expect the respective waveforms to deviate even further from their GR counterparts than the ones obtained ``at threshold'' and presented here. 
Once future research can establish that a given observational gravitational-wave catalogue statistically disfavors a beyond-GR waveform at threshold, we thus expect a respective lower bound on the value of~$m_2/M_\text{Pl}$.
While this will require dedicated statistical analysis, the present letter provides, for the first time, the means to obtain such bounds.
\\

%%%%%%%%%%%%%%%%%%%%%%%%%%%%%%%%%%%%%%%%%%%%%%%%%%%%%%%%%
%%%%%%%%%%%%%%%%%%%%%%%%%%%%%%%%%%%%%%%%%%%%%%%%%%%%%%%%%
\noindent{\textbf{Inspiral.}}
%%%%%%%%%%%%%%%%%%%%%%%%%%%%%%%%%%%%%%%%%%%%%%%%%%%%%%%%%
%%%%%%%%%%%%%%%%%%%%%%%%%%%%%%%%%%%%%%%%%%%%%%%%%%%%%%%%%
With quadratic curvature corrections present, the smaller black hole transitions to a non-GR black hole during inspiral. This causes the inspiral to occur faster and with higher inspiral frequency than in the respective GR case. As discussed above, we expect that the transition may lead to orbital eccentricity. It has been argued that gravitational radiation during a sufficiently long inspiral will circularise most astrophysically relevant binaries~\cite{PhysRev.136.B1224}.
Obtaining a quasi-circular beyond-GR system will require the implementation of eccentricity reduction algorithms~\cite{Pfeiffer:2007yz,Buonanno:2010yk}.
More generally, it remains desirable to directly construct consistent initial data in the non-Kerr branch such that the binary parameters can be set directly. 
It would also be of great interest to obtain analytic approximations for the inspiral waveform, including higher-curvature corrections in the Post-Newtonian (PN) and/or effective one-body (EOB) formalism, see, e.g.,~\cite{Damour:2009zoi} for review.
\\

%%%%%%%%%%%%%%%%%%%%%%%%%%%%%%%%%%%%%%%%%%%%%%%%%%%%%%%%%
%%%%%%%%%%%%%%%%%%%%%%%%%%%%%%%%%%%%%%%%%%%%%%%%%%%%%%%%%
\noindent{\textbf{Merger.}}
%%%%%%%%%%%%%%%%%%%%%%%%%%%%%%%%%%%%%%%%%%%%%%%%%%%%%%%%%
%%%%%%%%%%%%%%%%%%%%%%%%%%%%%%%%%%%%%%%%%%%%%%%%%%%%%%%%%
After the transition and throughout the merger, the spacetime deviates significantly from GR. We have verified this by tracking the average Ricci scalar curvature $\langle\mathcal{R}\rangle$ throughout the evolution and present respective plots in the supplementary material. 
We find that the plunge occurs more rapidly and the peak amplitude increases in comparison to the GR waveform, cf.~\cref{fig:waveforms}. While the qualitative merger dynamics remains similar to GR, the quantitive waveforms show a distinct mismatch. For the unequal-mass binary, we find that, after the merger, the Ricci scalar curvature decreases to zero, indicating that the merger remnant is a Kerr black hole. For the equal-mass binary, we find that the Ricci scalar curvature remains large, even post-merger, indicating a non-Kerr merger remnant.
\\

\begin{figure}[!t]
    \centering
    \includegraphics[trim={0cm 0cm 0cm 0cm},clip,width=0.48\linewidth]{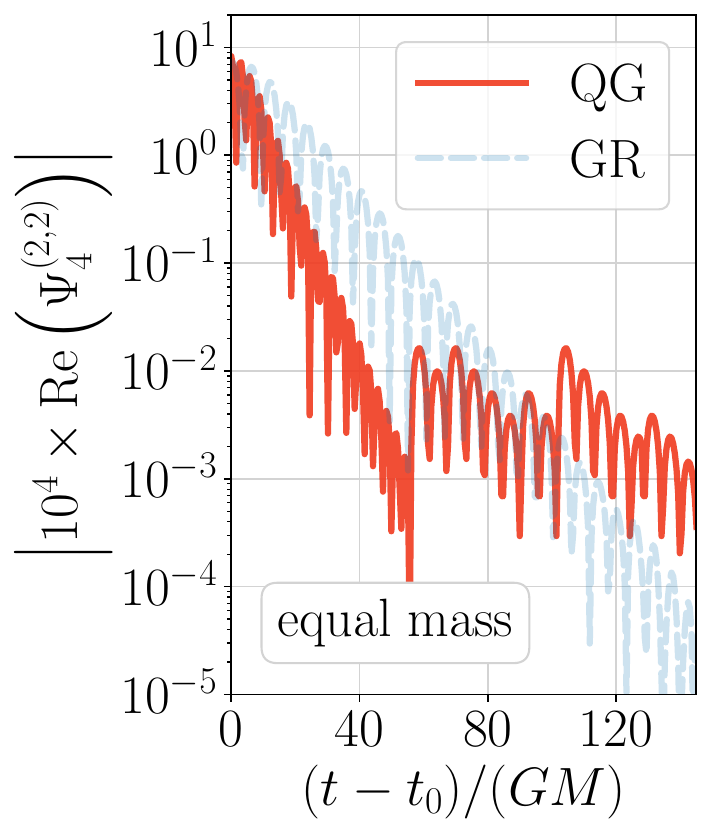}
    \includegraphics[trim={0cm 0cm 0cm 0cm},clip,width=0.48\linewidth]{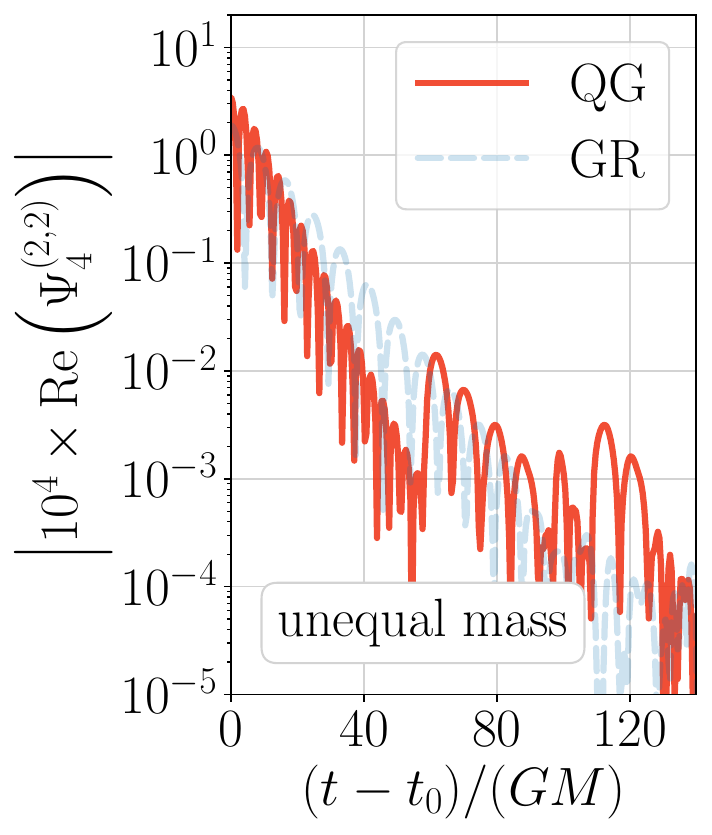}
    \caption{
    \label{fig:ringdown}
    Detailed view of the ringdown after the equal-mass (left) and unequal-mass (right) binary merger, cf.~\cref{fig:waveforms}. 
    }
\end{figure}

%%%%%%%%%%%%%%%%%%%%%%%%%%%%%%%%%%%%%%%%%%%%%%%%%%%%%%%%%
%%%%%%%%%%%%%%%%%%%%%%%%%%%%%%%%%%%%%%%%%%%%%%%%%%%%%%%%%
\noindent{\textbf{Ringdown.}}
%%%%%%%%%%%%%%%%%%%%%%%%%%%%%%%%%%%%%%%%%%%%%%%%%%%%%%%%%
%%%%%%%%%%%%%%%%%%%%%%%%%%%%%%%%%%%%%%%%%%%%%%%%%%%%%%%%%
As for GR, the waveforms suggest that the spacetime rings down to a single apparently stable, or at least longlived, black hole, see~\cref{fig:ringdown} for a logarithmic comparison of the post-merger waveforms in~GR and in~QG.
In GR, the waveform exhibits a single characteristic frequency, corresponding to the dominant quasi-normal mode of the final Kerr black hole. In QG, the post-merger waveform exhibits several distinct ringdown phases, dominated by different characteristic frequencies. Both, for the equal-mass and the unequal-mass system one of these frequencies seems to approximately match the dominant quasi-normal mode of the GR case. We expect that the other frequencies are associated with ringdown behaviour of the two massive modes.
A detailed study of the final state and its quasi-normal modes will appear in a separate publication.
\\

%%%%%%%%%%%%%%%%%%%%%%%%%%%%%%%%%%%%%%%%%%%%%%%%%%%%%%%%%
%%%%%%%%%%%%%%%%%%%%%%%%%%%%%%%%%%%%%%%%%%%%%%%%%%%%%%%%%
\noindent{\textbf{Discussion.}}
%%%%%%%%%%%%%%%%%%%%%%%%%%%%%%%%%%%%%%%%%%%%%%%%%%%%%%%%%
%%%%%%%%%%%%%%%%%%%%%%%%%%%%%%%%%%%%%%%%%%%%%%%%%%%%%%%%%
We have obtained the first complete wave-form predictions of black-hole binaries deviating from GR due to quadratic curvature corrections. Apart from the standard numerical discretization of hyperbolic PDEs (which converges to the continuum field theory at the expected rate), our results are fully nonlinear and obtained without any additional approximation. The obtained waveforms deviate from their GR counterpart and we, therefore, expect that future statistical comparison and parameter inference will provide observational strong-field constraints on the presence of said corrections.

In addition to the massless graviton of GR, the field equations of QG propagate a massive tensor field with mass~$m_2$ and a massive scalar with mass~$m_0$. We have fixed~$m_2$ ``at threshold'', i.e., such that it maximizes a well-known linear instability rate around Schwarzschild spacetime, see~\cref{fig:transitions} and~\cite{Brito:2013wya,Lu:2017kzi,Collingbourne:2020sfy,Held:2022abx}.
While the required more complete coverage of the $(m_0, m_2)$ parameter space will be presented in upcoming work, our results already suggest that a statistical comparison of observational data with binary simulations ``at threshold'' can provide a lower mass bound in the range of $GM_i,m_2 \gtrsim 0.2$ set by mass scale $M_i$ of the lightest observed binary companion. A related bound on $m_0$ will likely be weaker and may depend on the former bound.

The QG parameters relate to effective field theory (EFT) corrections to GR. To be precise, we refer to a covariant EFT of gravity which assumes that (i) the only light degree of freedom is the massless metric, (ii) the EFT expands in powers of curvature, and (iii) all dimensionless EFT couplings have natural values. Within this EFT, the additional massive modes are considered to be a truncation artefact. Here, we focus on binary systems involving black-hole solutions which do not continuously deform to Kerr black holes in the limit of heavy beyond-GR masses. This suggests that the presented simulations are not within the regime of validity of the above EFT. 
To explore deviations within the EFT, simulations at $GM\,m_2\approx0.44$ will be most interesting because, in this case, at least the two spherically-symmetric branches of black-hole solutions are perturbatively close to each other. We will explore this regime, as well as the interplay of fiducial modes, field redefinitions, regime of validity, and nonlinear dynamical evolution, in a separate publication. 
The presented evolution scheme and extensions thereof to higher order in curvature (see~\cite{Figueras:2024bba} for well-posedness at higher order) provide the means for ``numerical evolution without approximation'' and thus the basis for a faithful comparison.
If indeed all effects of quadratic-curvature terms can be removed from physical observables in vacuum spacetimes, we expect that the tightest gravitational-wave constraints on the scale of new physics within the above gravitational EFT will then arise from quadratic-curvature corrections in the presence of matter. Future work on simulations of neutron-star mergers in Quadratic Gravity, see~\cite{Cayuso:2023dei} for related simulations of spherically symmetric gravitational collapse, is thus of great phenomenological interest. 

Alternatively, the action of QG may be interpreted as a fundamental theory of gravity. In this case, the extra massive modes are physical. 
This interpretation is typically dismissed due to the opposite signs of the kinetic terms for the massless and the massive tensor modes and the related expectation that this causes a catastrophic instability~\cite{Ostrogradsky:1850fid}. We caution that this expectation derives from physical arguments extending results for classical point-particle systems and, even for the latter, proven stable counterexamples exist~\cite{Deffayet:2021nnt,Deffayet:2023wdg}. If anything, our simulations suggest that the classical continuum field theory does not develop a catastrophic instability, at least for the specified initial data and for the given evolution time.
\\

\paragraph*{Acknowledgements.}
We thank Miguel Bezares, Ramiro Cayuso, Katy Clough, Will East, Pau Figueras, \'Aron Kov\'acs, Nils Siemonsen, and David Van Komen for discussion.
HL is supported by the LANL LDRD grant 20220087DR. This work used resources provided by the LANL Darwin testbed. Darwin is a research testbed/heterogeneous cluster funded by the Computational Systems and Software Environments subprogram of ASC program. LANL is operated by Triad National Security, LLC, for the National Nuclear Security Administration of the U.S.DOE  (Contract No. 89233218CNA000001). This work is authorized for unlimited release under LA-UR-24-24999.

%%%%%%%%%%%%%%%%%%%%%%%%%%%%%%%%%%%%%%%%%%%%%%%%%%%%%%%%%
%%%%%%%%%%%%%%%%%%%%%%%%%%%%%%%%%%%%%%%%%%%%%%%%%%%%%%%%%
%%%%%%%%%%%%%%%%%%%%%%%%%%%%%%%%%%%%%%%%%%%%%%%%%%%%%%%%%
%\clearpage
\bibliography{References}
%%%%%%%%%%%%%%%%%%%%%%%%%%%%%%%%%%%%%%%%%%%%%%%%%%%%%%%%%
%%%%%%%%%%%%%%%%%%%%%%%%%%%%%%%%%%%%%%%%%%%%%%%%%%%%%%%%%
%%%%%%%%%%%%%%%%%%%%%%%%%%%%%%%%%%%%%%%%%%%%%%%%%%%%%%%%%

%%%%%%%%%%%%%%%%%%%%%%%%%%%%%%%%%%%%%%%%%%%%%%%%%%%%%%%%%
%%%%%%%%%%%%%%%%%%%%%%%%%%%%%%%%%%%%%%%%%%%%%%%%%%%%%%%%%
%%%%%%%%%%%%%%%%%%%%%%%%%%%%%%%%%%%%%%%%%%%%%%%%%%%%%%%%%
\clearpage
\appendix
%%%%%%%%%%%%%%%%%%%%%%%%%%%%%%%%%%%%%%%%%%%%%%%%%%%%%%%%%
%%%%%%%%%%%%%%%%%%%%%%%%%%%%%%%%%%%%%%%%%%%%%%%%%%%%%%%%%
%%%%%%%%%%%%%%%%%%%%%%%%%%%%%%%%%%%%%%%%%%%%%%%%%%%%%%%%%

%\begin{widetext}
\begin{center}
{\large Supplementary Material}
\end{center}
%\end{widetext}

%%%%%%%%%%%%%%%%%%%%%%%%%%%%%%%%%%%%%%%%%%%%%%%%%%%%%%%%%
%%%%%%%%%%%%%%%%%%%%%%%%%%%%%%%%%%%%%%%%%%%%%%%%%%%%%%%%%
\section{Numerical implementation}
\label{app:nr}
%%%%%%%%%%%%%%%%%%%%%%%%%%%%%%%%%%%%%%%%%%%%%%%%%%%%%%%%%
%%%%%%%%%%%%%%%%%%%%%%%%%%%%%%%%%%%%%%%%%%%%%%%%%%%%%%%%%

\subsection{Numerical evolution with Dendro-GR}
\label{app:nr:code}

We employ \texttt{Dendro-GR} to evolve our system. \texttt{Dendro-GR} integrates a parallel octree-refined adaptive mesh with a wavelet-based adaptive multiresolution approach, allowing for efficient and accurate numerical computations. We have developed an additional Quadratic-Gravity module on top of this framework. \texttt{Dendro-GR} is an open-source tool\footnote{For details on building the Quadratic-Gravity module, see the \texttt{GitHub} repository \url{https://github.com/lanl/Dendro-GRCA} and refer to the \texttt{README.md} file.}. For spatial derivatives, we utilize a fourth-order finite-difference scheme, while time evolution is performed using a fourth-order Runge-Kutta method. In the numerical simulations presented in this work, we use 15 levels of wavelet refinement and the finest resolution is $\Delta x_\textrm{min} \approx 0.004 M$ with a Courant–Friedrichs–Lewy (CFL) condition~\cite{Courant:1967} of 0.25. Further details on the convergence properties of our numerical approach can be found in~\cite{Held:2022abx,Held:2023aap}

\subsection{Diagnostics}
\label{app:nr:diag}

To compute the apparent horizon, we employ well-established numerical techniques as outlined in~\cite{Thornburg:2006zb}. Specifically, we utilize the \texttt{AHFinderDirect} thorn, a component of the \texttt{EinsteinToolkit}~\cite{Loffler2011ay}.

To obtain the gravitational wave, we calculate the Weyl scalar $\Psi_4$. In our implementation within \texttt{Dendro-GR}, we adhere to the conventions established in~\cite{Brugmann2007}. The computed $\Psi_4$ is then decomposed into multipolar components using spin-weighted spherical harmonics $_{s}Y_{lm}$ with spin-weighted $s=-2$. This evaluation is performed at a chosen extraction radius $R$, i.e., at $R=100M$ in our case. 

To track the departure from the Ricci-flat GR manifold, we obtain the evolution of the Ricci scalar curvature $\langle\mathcal{R}\rangle\equiv\langle\mathcal{R}\rangle_\zeta$, averaged within each spatial slice, where the spatial average is obtained in a cube with $x,\,y,\,z\in[-\zeta,+\zeta]$ and $\zeta$ extends across the full computational domain. For the single black-hole runs, $\zeta=250\,GM$. For the binary black-hole runs, $\zeta=300\,GM$.

%%%%%%%%%%%%%%%%%%%%%%%%%%%%%%%%%%%%%%%%%%%%%%%%%%%%%%%%%
%%%%%%%%%%%%%%%%%%%%%%%%%%%%%%%%%%%%%%%%%%%%%%%%%%%%%%%%%
\section{Black-hole transitions}
\label{app:transitions}
%%%%%%%%%%%%%%%%%%%%%%%%%%%%%%%%%%%%%%%%%%%%%%%%%%%%%%%%%
%%%%%%%%%%%%%%%%%%%%%%%%%%%%%%%%%%%%%%%%%%%%%%%%%%%%%%%%%

%
\begin{figure}[!t]
    \centering
    \includegraphics[trim={0cm 0cm 0cm 0cm},clip,width=\linewidth]{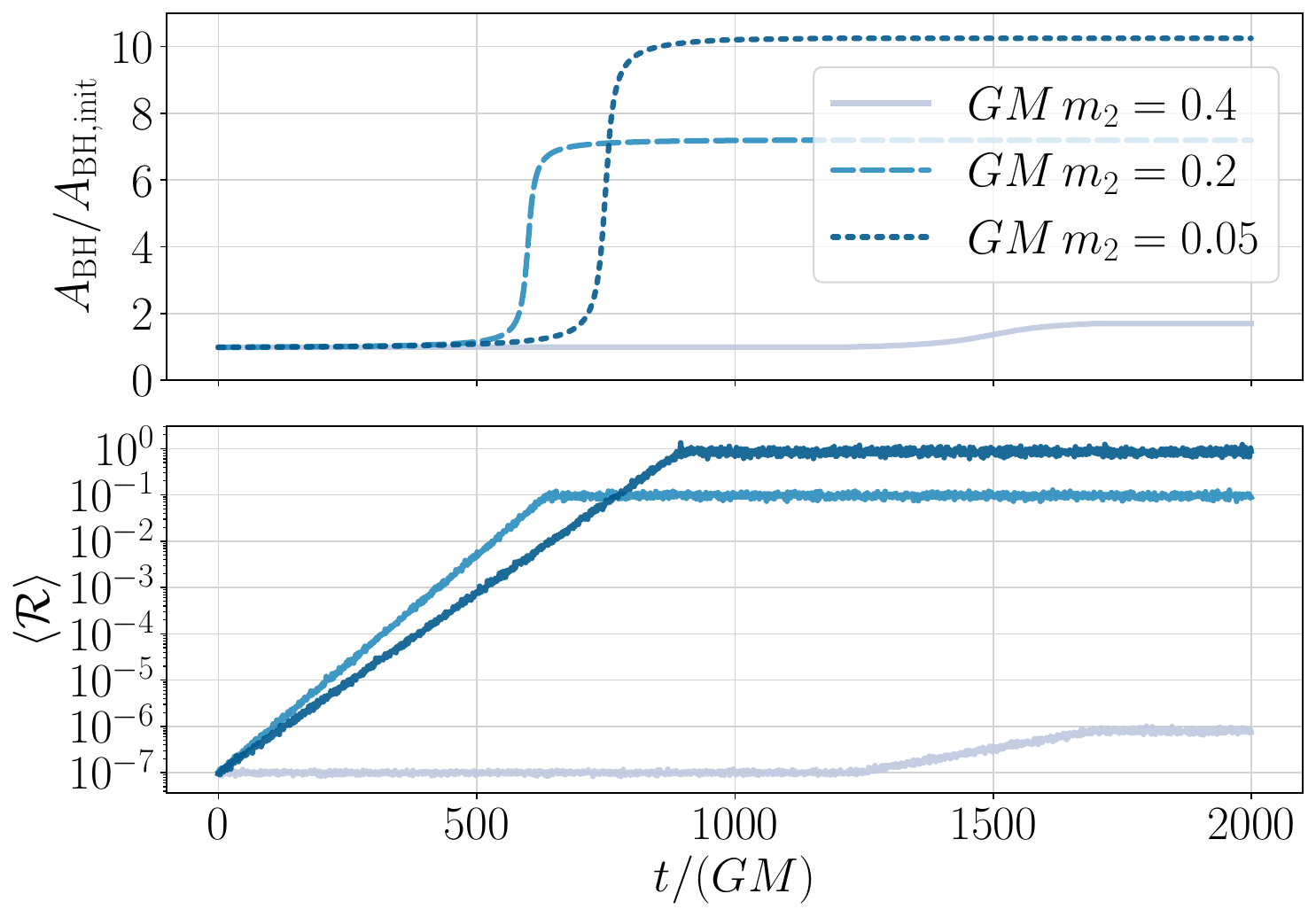}
    \caption{
    \label{fig:transitions_curvature}
    Upper panel: As in~\cref{fig:transitions} and provided here again for direct comparison with the lower panel.
    Lower panel: Evolution of the spatial average of Ricci scalar curvature $\langle\mathcal{R}\rangle$. 
    While the apparent horizon grows, the spacetime develops non-vanishing Ricci scalar curvature.
    }
\end{figure}
\begin{figure}[!t]
    \centering
    \includegraphics[trim={0cm 0cm 0cm 0cm},clip,width=\linewidth]{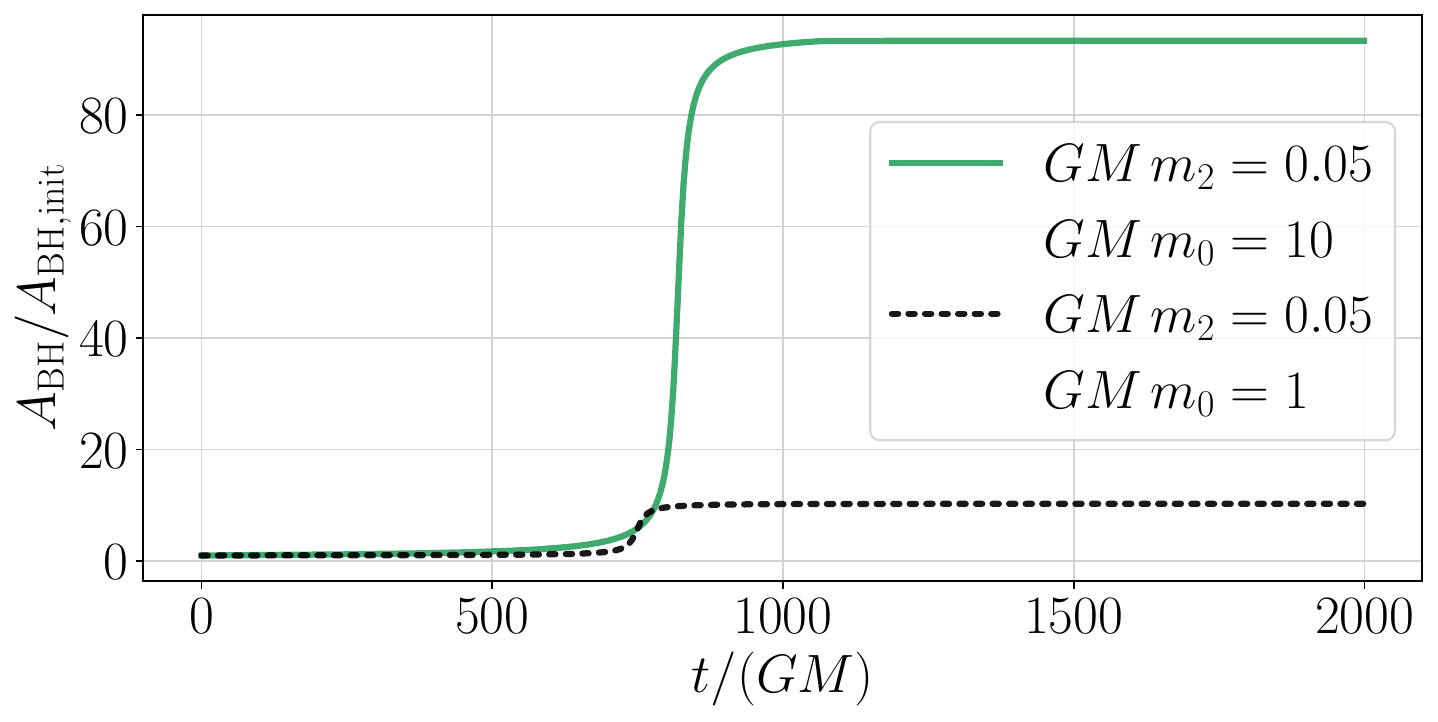}
    \caption{
    \label{fig:transitions_decoupling_scalar_mode}
    As in~\cref{fig:transitions}, we show transitions of Schwarzschild to non-Schwarzschild black holes as indicated by a growth of the apparent horizon~$A_\text{BH}$. The dashed line ($GM\,m_2=0.05$ and $GM\,m_0=1$) corresponds to the dashed line in~\cref{fig:transitions}. The other case ($GM\,m_2=0.05$ and $GM\,m_0=10$), cf.~solid line, demonstrates that, in the limit $m_0/M_\text{Pl}\rightarrow \infty$, our results seem to agree with those of~\cite[Fig.~2]{East:2023nsk}.
    }
\end{figure}

Here, we provide further detail on the single black-hole transitions, including a comparison to results in~\cite{East:2023nsk} which are equivalent to the $m_0/M_\text{Pl}\rightarrow \infty$ limit.

In~\cref{fig:transitions_curvature}, we show the evolution of spatially averaged Ricci scalar curvature $\langle\mathcal{R}\rangle$ (see~\cref{app:nr:diag}) alongside a duplicated plot of the black-hole transitions discussed in the main text, cf.~\cref{fig:transitions}. This juxtaposition demonstrates that the Ricci scalar curvature grows during the transitions and then saturates as the black hole, i.e., the apparent horizon, converges to its final non-Kerr form.

Second, we are interested in a comparison with the only other nonlinear study of Quadratic Gravity~\cite{East:2023nsk} where similar black-hole transitions have been obtained. We highlight several important distinctions between our evolution code and the one in~\cite{East:2023nsk}: While we treat the GR sector in BSSN form, \cite{East:2023nsk} uses the harmonic-gauge formalism. Further, while we evolve the full (3+1) system without any symmetry assumptions, \cite{East:2023nsk} employs a symmetry reduction and only evolves the axially symmetric sector. Finally, while we work with finite scalar mass~$m_0$, the evolution in \cite{East:2023nsk} does not include the massive scalar mode and thus corresponds to the respective infinite-mass decoupling limit. 
To compare to the transitions in~\cite[Fig.~2]{East:2023nsk} obtained in the $m_0/M_\text{Pl}\rightarrow \infty$ limit we thus need to approach transitions at large scalar mass $m_0$. In~\cref{fig:transitions_decoupling_scalar_mode} we show respective transitions for $GM\,m_2=0.05$ and $GM\,m_0=1$ (as in the main text) as well as $GM\,m_0=10$, i.e., approaching the limit of heavy scalar mass. Indeed, we find that the respective transition seems to converge to the respective one obtained in~\cite[Fig.~2]{East:2023nsk}. We note that to compare these results one needs to convert between the apparent horizon and the local Christodoulou mass (or, equivalently, horizon radius).

%%%%%%%%%%%%%%%%%%%%%%%%%%%%%%%%%%%%%%%%%%%%%%%%%%%%%%%%%
%%%%%%%%%%%%%%%%%%%%%%%%%%%%%%%%%%%%%%%%%%%%%%%%%%%%%%%%%
\section{Full gravitational waveforms}
\label{app:full-waveforms}
%%%%%%%%%%%%%%%%%%%%%%%%%%%%%%%%%%%%%%%%%%%%%%%%%%%%%%%%%
%%%%%%%%%%%%%%%%%%%%%%%%%%%%%%%%%%%%%%%%%%%%%%%%%%%%%%%%%

\begin{figure*}[h]
    \centering
    \includegraphics[trim={0cm 0cm 0cm 0cm},clip,width=0.49\linewidth]{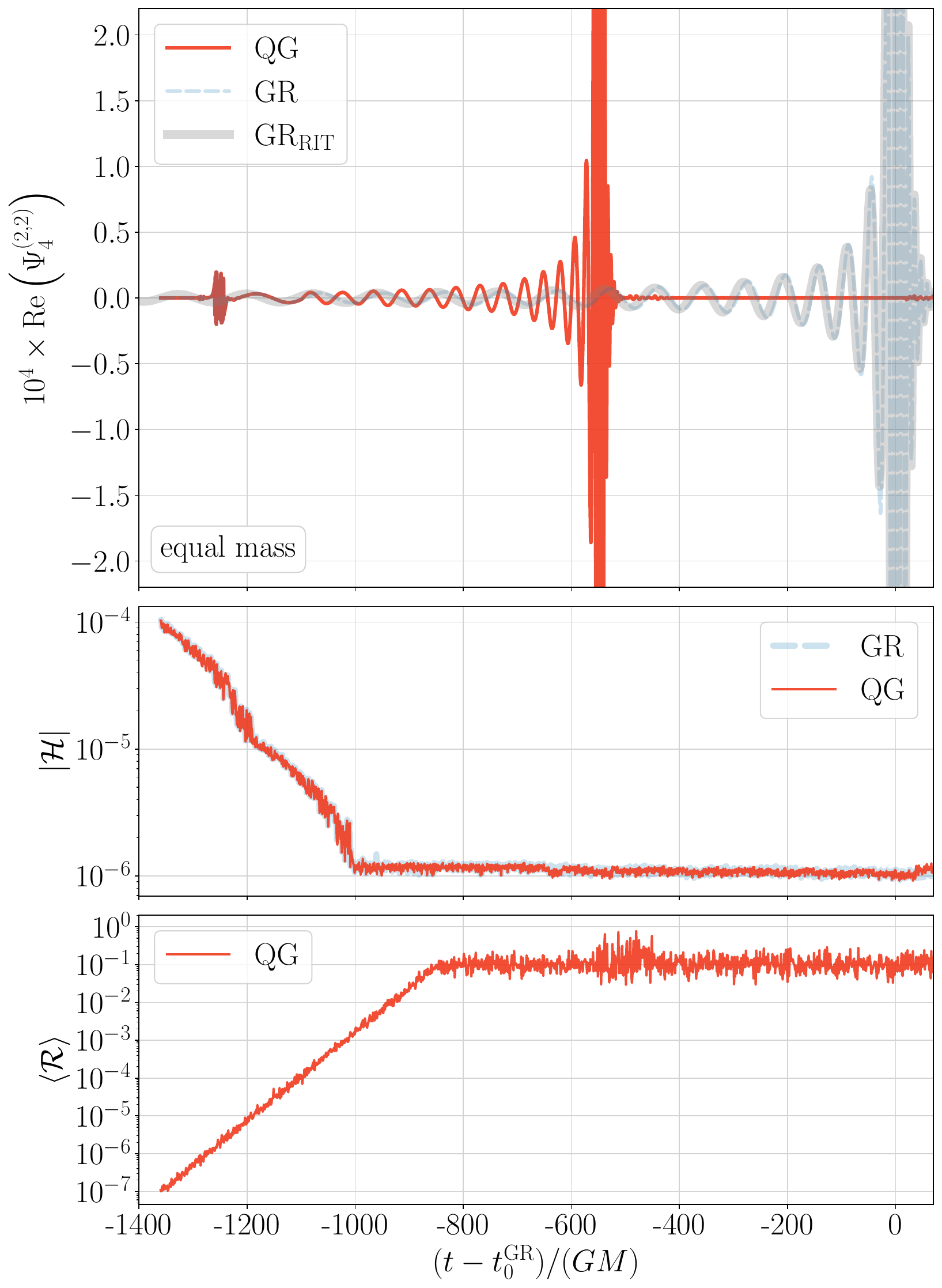}
    \includegraphics[trim={0cm 0cm 0cm 0cm},clip,width=0.49\linewidth]{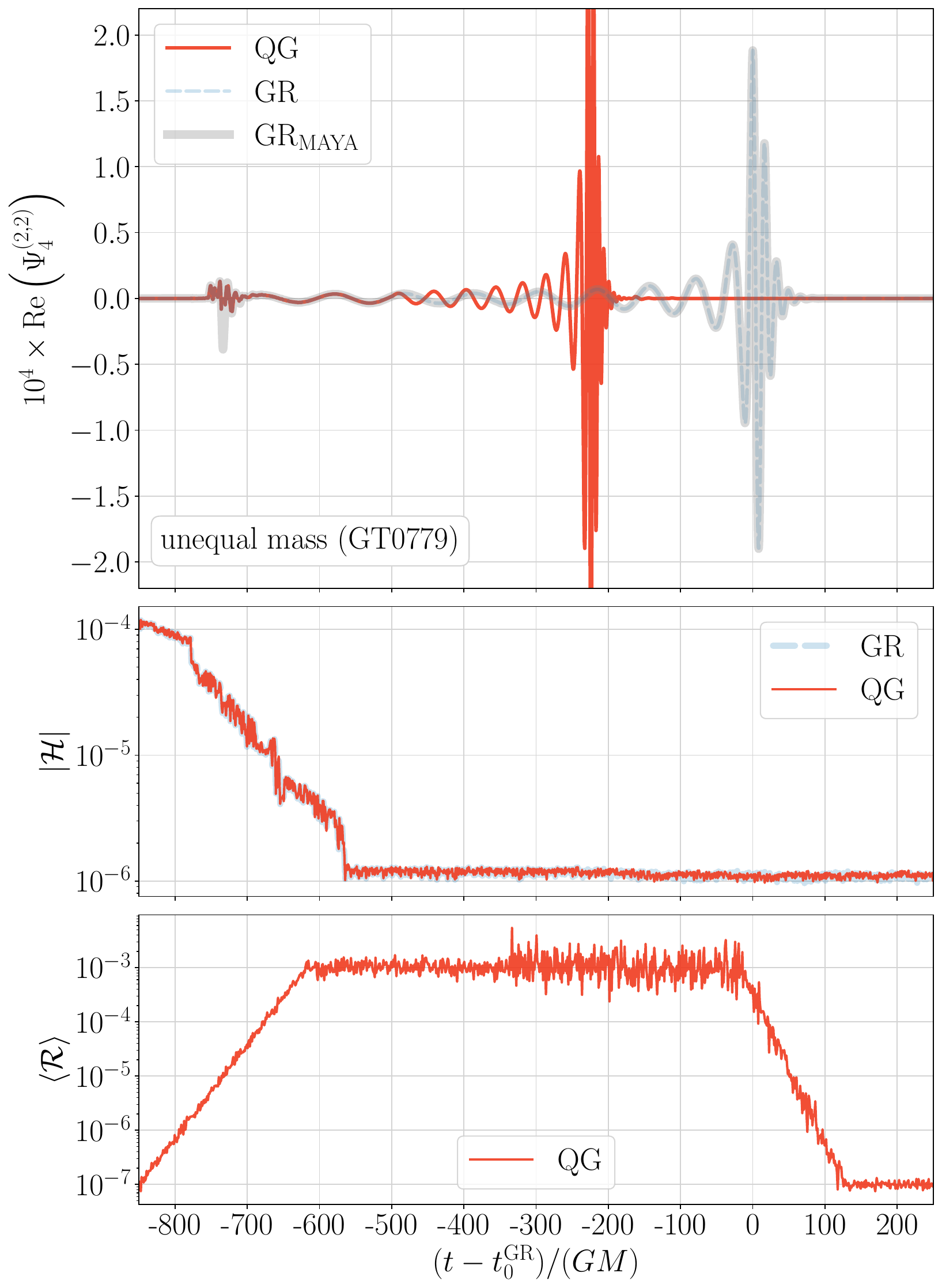}
    \caption{
    \label{fig:waveforms_full}
    Upper panels: Gravitational waveforms for an equal-mass (left) and an unequal-mass (right) binary merger, see~\cref{tab:binary-params} for the initial binary parameters and QG masses. In contrast to the main text, all waveforms are adjusted to the merger time of GR $t_0^\text{GR}$ (in units of the total mass $M$) and, moreover, we here include the early-inspiral portion of the simulations. This includes junk radiation occurs (around $t-t_0^\text{GR}=-1250\,GM$ in the equal and around $t-t_0^\text{GR}=-750\,GM$ in the unequal mass case) and a subsequent regime in which the individual black holes transition from Kerr to non-Kerr black holes.
    Center panels: Respective violations of the Hamiltonian constraint $|\mathcal{H}|$ (averaged within each time slice) which occur due to the discretisation. 
    Bottom panels:
    Spatial average $\langle\mathcal{R}\rangle$ of the Ricci scalar curvature. 
    }
\end{figure*}

For completeness, we present the full gravitational waveforms, including unphysical junk radiation, i.e., radiation which occurs due to non-equilibrium initial data, and the unphysical ``early inspiral'', i.e., the portion of the waveform during which respective transitions from Kerr to non-Kerr black holes occur. The respective waveforms and their comparison to GR are shown in the upper panels of~\cref{fig:waveforms_full}. 
As a crosscheck of the employed \texttt{Dendro-GR} code, we also overlay the respective waveforms obtained from the \texttt{RIT}~\cite{Healy:2022wdn} and the \texttt{MAYA}~\cite{Ferguson:2023vta}  catalog (see $\text{GR}_\text{RIT}$ and $\text{GR}_\text{MAYA}$ in the upper panels of~\cref{fig:waveforms_full}). We attribute small changes in comparison to the \texttt{RIT} waveform to their extrapolation of the waveform to asymptotic infinity.
\\

In contrast to the figures in the main text (see~\cref{fig:waveforms}), the waveform plots in the upper panels of~\cref{fig:waveforms_full} are not aligned to merger time but rather in the early inspiral regime, i.e., because we use the same initial data for the GR and the QG simulation. Hence, in the early inspiral, i.e., before the black-hole transitions occur (up to $t/(GM)<-1200$ for the equal mass and up to $t/(GM)<-500$ for the unequal mass), the QG and GR waveforms are almost indistinguishable. 
Showing the full waveforms, and aligning them in the early inspiral, corresponds to the (unphysical) assumption that the transition occurs at this particular time during the inspriral. Given that the respective Kerr BH is unstable in isolation, we consider this an unphysical or, at least, highly unlikely scenario.
\\

In the center panels of~\cref{fig:waveforms_full}, we show the Hamiltonian constraint as a function of evolution time. The smallness of the constraint violations suggests that the obtained solutions approximate the continuum field theory well. Convergence rates of the Quadratic Gravity module of \texttt{Dendro-GR} have been verified in~\cite{Held:2023aap}.
\\

Finally, in the bottom panels of~\cref{fig:waveforms_full}, we show the spatial average $\langle\mathcal{R}\rangle$ of the Ricci scalar curvature as a function of evolution time. Since the GR cases remain Ricci flat throughout evolution, we only show the Ricci scalar curvature for the QG cases. In both binaries, the Ricci scalar curvature grows during the early inspiral phase, indicating a transition of the individual Kerr black hole(s) to non-Kerr black holes as in the single black-hole studies, cf.~\cref{app:transitions}. We note that it is not the total mass ratio $m_2/M$ but rather the individual black-hole mass ratios, i.e., $m_2/M_1$ and $m_2/M_2$ which determine the onset and behaviour of these transitions. We have chosen $m_2$ such as to minimise the transition time of the smaller-mass black hole, i.e., both black holes in the equal-mass system and the lighter black hole with mass $M_2$ in the unequal-mass binary. Throughout the subsequent inspiral, the merger, and the ringdown phase, the spacetime exhibits significant Ricci scalar curvature, i.e., is no longer on the Ricci-flat GR manifold. Comparing the two simulated binaries, we find that the post-merger behaviour is distinct: For the equal-mass system, the average Ricci scalar curvature remains at its non-vanishing value. In contrast, for the unequal-mass system, the average Ricci scalar curvature decays. We conclude that the unequal mass merger results in a Kerr remnant black hole while the equal mass merger does not.

\end{document}